\documentclass{PoS}
\usepackage{graphicx}
\newcommand{\ba}{\begin{eqnarray}}
\newcommand{\ea}{\end{eqnarray}}
\newcommand{\be} {\begin{equation}}
\newcommand{\ee} {\end{equation}}

\newcommand{\order}{{\cal O}}

\title{$B_s$ and $B_d$ mixing in full lattice QCD 
using NRQCD $b$ quarks}

\ShortTitle{$B_s$ and $B_d$ mixing in full lattice QCD 
using NRQCD $b$ quarks}

\author{Elvira G\'amiz\\
        Department of Physics, University of Illinois, Urbana, IL 61801, USA\\
        E-mail: \email{megamiz@uiuc.edu}}

\author{\speaker{Christine T.H.~Davies}\\
        Department of Physics \&
               Astronomy, University of Glasgow, Glasgow, G12 8QQ, UK\\
        }

\author{G.~Peter Lepage\\
        LEPP, Cornell University, Ithaca, NY 14853, USA\\
       }

\author{Junko Shigemitsu\\
Physics Department, The Ohio State
        University, Columbus, OH 43210, USA\\
        }

\author{Matthew Wingate\\
Centre for Mathematical Sciences, University of Cambridge, 
Cambridge, CB3 0WA, UK\\
        }

\author{HPQCD Collaboration}

\abstract{  
We give a progress report on studies of $B_s$ and $B_d$ mixing with 
valence NRQCD $b$ quarks and asqtad light quarks on the MILC 
configurations including the effect of 2+1 flavours of sea quarks. We 
explore methods for reducing statistical and systematic errors in 
the ratio $\xi = f_{B_s}\sqrt{B_{B_s}}/f_{B_d}\sqrt{B_{B_d}}$.
}

\FullConference{The XXV International Symposium on Lattice Field Theory\\
		 July 30-4 August 2007\\
		 Regensburg, Germany}

\begin{document}

\section{Introduction}

%  The inputs in Figure \ref{unitriangle}, corresponding to 
%lattice calculations performed by the HPQCD collaboration, are taken from 
%\cite{BKHPQCD} ($\vert\varepsilon_K\vert$), \cite{fplus,HFAG05} 
%($\vert V_{ub}\vert$) and \cite{fBpaper,BBJLQCD} ($\vert \frac{V_{td}}{V_{ts}}
%\vert$).

%\begin{figure}
%    \begin{center}              
%            \includegraphics[width=12cm,height=8cm]{ckm.eps} 
%        \end{center}    
%\caption{Unitarity triangle with inputs for $\vert \varepsilon_K\vert$ 
%taken from HPQCD lattice calculations.\label{unitriangle}}
%\end{figure}

The precise determination of the elements of the 
Cabibbo-Kobayashi-Maskawa (CKM) matrix can impose important constraints 
on physics beyond the Standard Model (SM). One combination of 
CKM matrix elements that plays a relevant role in this analysis 
is  $\left\vert\frac{V_{td}}{V_{ts}}\right\vert$, which is related to 
$B_0-\bar B_0$ mixing. 

In particular, this combination of CKM matrix elements can be 
extracted from the precisely experimentally measured quantities 
$\Delta M_s$ and $\Delta M_d$, which are the mass differences between 
the heavy and light mass eigenstates in the 
$B^0_s-\bar B^0_s$ and $B^0_d-\bar B^0_d$ systems respectively. 
The relation is given by
\begin{equation}\label{CKMratio}
\left\vert\frac{V_{td}}{V_{ts}}\right\vert = 
\frac{f_{B_s}\sqrt{B_{B_s}}}{f_{B_d}\sqrt{B_{B_d}}} 
\sqrt{\frac{\Delta M_d M_{B_s}}{\Delta M_s M_{B_d}}}\, .
\end{equation}

The masses $M_{B_s}$ and $M_{B_d}$, and the corresponding mass differences 
are known experimentally with very high precision \cite{MsMdmeasurements}. 
For the ratio $\xi= \frac{f_{B_s}\sqrt{B_{B_s}}}{f_{B_d}\sqrt{B_{B_d}}}$, 
however, an accurate and consistent lattice calculation 
that fully incorporates vacuum polarization effects is not yet available. 
Our goal is to 
perform such a calculation and reduce the theoretical errors in the 
ratio $\xi$ to a few percent. This will provide us with a high 
precision determination of  the CKM ratio in (\ref{CKMratio}).

The products of $B_0$ decay constants and bag parameters in (\ref{CKMratio}) 
are determined by  matrix elements between $B_0$ and 
$\bar B_0$ of the four-fermion operators appearing in the effective 
hamiltonian that describes $\Delta B=2$ processes. The non-perturbative 
inputs for the calculation of ${\Delta \Gamma_s}$ and ${\Delta \Gamma_d}$ 
(with $\Delta \Gamma$ the width difference between the light and heavy 
mass eigenstate) are also given by this kind of hadronic matrix elements. 
For completeness, we are studying all the matrix elements needed to make 
theoretical predictions for ${\Delta M_s}$, ${\Delta M_d}$, 
${\Delta \Gamma_s}$ and ${\Delta \Gamma_d}$.

\section{Simulation details and milestones in the calculation}

The four-fermion operators whose matrix element between $B_0$ and 
$\bar B_0$ are needed to make a complete study of $B_0^s$ 
and $B_0^d$ mixing in the SM are
\ba\label{operators}
&OL^q \equiv \left[\bar b^i q^i\right]_{V-A} 
\left[\bar b^j q^j\right]_{V-A}\,;\quad\quad\quad
OS^q \equiv \left[\bar b^i q^i\right]_{S-P} 
\left[\bar b^j q^j\right]_{S-P}\,;&\nonumber\\
&O3^q \equiv \left[\bar b^i q^j\right]_{S-P} 
\left[\bar b^j q^i\right]_{S-P}\,;&\nonumber\\
&  OLj1^q   
\equiv \frac{1}{2M} \left\{ [\vec{\nabla}\overline{b^i}
 \cdot \vec{\gamma} \, q^i]_{V-A} [\overline{b^j} \,
q^j]_{V-A} + 
  [\overline{b^i} \, q^i]_{V-A} [\vec{\nabla}\overline{b^j}
 \cdot \vec{\gamma} \, q^j]_{V-A}  \right\} &\, ;
\ea
with $q$ being a strange or a down quark, and $i$, $j$ colour indices. 
The last operator, as well as similar $1/M$ corrections $OSj1^q$ and 
$O3j1^q$ for the $OS^q$ and $O3^q$ operators, are required 
at $\order\left(\Lambda_{QCD}/M\right)$. 

The continuum matrix elements $\langle OX \rangle(\mu)^{\overline{MS}}\equiv 
\langle\bar B_0^q\vert OX\vert B_0^q\rangle^{\overline{MS}}(\mu)$ of 
the operators $OX=OL^q$, $OS^q$, $O3^q$ entering in the SM formulae,  
are related to those evaluated via lattice 
simulations by a perturbative one-loop matching relation through 
$\order\left(\alpha_s\right)$, $\order\left(\Lambda_{QCD}/M\right)$ 
and $\order\left(\alpha_s/(aM)\right)$. The matching relations mix, 
already in the continuum, the four-fermion operators in 
(\ref{operators}) -see \cite{Bspaper,Bsproc} for the explicit expressions.

The bare hadronic matrix elements are obtained by numerically evaluating 
the three-point and two-point correlation functions 
\ba\label{corrdef}
&C^{(4f)}(t_1,t_2)= \displaystyle\sum_{\vec{x}_1,\vec{x}_2}\langle 0\vert 
\Phi_{\bar{B}_q}(\vec{x}_1,t_1)\left[\hat Q\right](0)
\Phi_{\bar{B}_q}{\dagger}(\vec{x}_2,-t_2)\vert 0\rangle&\nonumber\\
&C^{(B)}(t)= \displaystyle\sum_{\vec{x}}\langle 0\vert 
\Phi_{\bar{B}_{q}}(\vec{x},t)
\Phi_{\bar{B}_{q}}^{\dagger}(\vec{0},0)\vert 0\rangle&
\ea
with $\Phi_{\bar{B_q}}(\vec{x},t)=\bar b (\vec{x},t)\gamma_5
q(\vec{x},t)$ and $\hat O$ any of the four-fermion operators in 
(\ref{operators}). The simulations 
are performed on MILC configurations with $N_f=2+1$ sea quarks. 
The valence $b$ fields are described by the NRQCD action 
improved through  ${\order(1/M^2)}, \,{\order(a^2)}$ and leading 
relativistic ${\order(1/M^3)}$ \cite{NRQCD}, while the light valence 
(and sea) quarks are staggered asqtad fields \cite{Asqtadaction}. 
An improved gluon action is also used to further reduce discretization 
errors.

The action parameters are fixed via light and heavy-heavy simulations, 
in particular the valence $b$ and $s$ quark masses are tuned to give 
the physical values of the $\Upsilon$ and $K$ mesons. The different 
parameters in the simulations are collected in Table \ref{parameters}. 

\begin{center}
\begin{table}[t]
\begin{center}
\begin{tabular}{c c c c c c c}\hline\hline
$m_{light}^{sea}/m_s^{phys.}$ & Volume & $N_{confs}$ & $a(fm)$ 
& $a m_b$ & $m_q^{val.}/m_s^{phys.}$ & $N_{sources}$\\
\hline
0.5 & $20^3\times 64$ & 486 & 0.12 & 2.8 & 
\begin{tabular}{c} 1 \\ 0.5 \end{tabular} & 
\begin{tabular}{c} 1 \\ 2 \end{tabular}\\
0.25 & $20^3\times 64$ & 568  & 0.12 & 2.8 &
\begin{tabular}{c} 1 \\ 0.25 \end{tabular} & 
\begin{tabular}{c} 1 \\ 2 \end{tabular} \\
0.175 & $20^3\times 64$ & 402   & 0.12 & 2.8 & 1, 0.175  & 4\\
0.125 &  $24^3\times 64$ & 678    & 0.12 & 2.8 & 1, 0.125  & 4\\
\hline
0.2 &  $28^3\times 96$ & 563   & 0.09 & 1.95 & 1, 0.2  & 4\\
\hline
\hline
\end{tabular}
\end{center}
\caption{Simulation parameters for the coarse (first four sea masses) and 
fine lattices (last line) for $B_0^s$ ($m_q^{val.}/m_s^{phys.}=1$)  
and $B_0^d$ ($m_q^{val.}=m_d^{sea}$).\label{parameters}}
\end{table}
\end{center}

\subsection{Mixing parameter for $B_0^s$ mixing} 

Our work published in \cite{Bspaper} analyzes the $B^0_s$ mixing 
parameters for two ensembles of MILC configurations with 
$(m_u^{sea}=m_d^{sea})/m_s=0.25,0.50$ and $a=0.12fm$ (coarse lattice). 
This corresponds to the first two entries in Table \ref{parameters} 
with $m_q^{val.}/m_s^{phys.}=1$.

The results obtained for the mass and width differences when using 
these parameters in the SM expressions are 
\ba\label{masaresult}
\Delta M_s = 20.3(3.0)(0.8)ps^{-1}\quad{\rm and}\quad
\Delta \Gamma_s = 0.10(3)ps^{-1}\,, 
\ea
which agree with experimental results within errors. The first error 
in $\Delta M_s$, which is the dominant one, is from the lattice 
determination of $f_{B_s}^2 B_{B_s}$ through the definition
\ba
\langle OL \rangle ^{\overline{MS}}_{(\mu)}
\equiv
\langle \overline{B}_s | OL| B_s \rangle ^{\overline{MS}}_{(\mu)} 
\equiv \frac{8}{3}
f^2_{B_s} \, B_{B_s}(\mu)\,  M^2_{B_s}\, ,
\ea
and the second one is an estimate of the error from 
$\vert V_{ts}^*V_{tb}\vert$ and $\overline{m_t}$. This $15\%$ lattice error
is dominated by a $9\%$ statistics+fitting error and a $9\%$ uncertainty 
associated with higher order operator matching. The large statistical errors 
are due to the fact that the simultaneous fits of two-point 
and three-point functions are unstable and we need to constrain the 
two-point parameters using the values obtained in fits to only two-point 
correlators. 

    The stability of the fits can be improved by using 
smearing techniques that reduce the overlap with excited states. We checked   
that the statistical+fitting error can be reduced from $9\%$ down to 
as low as $2\%$ by smearing the heavy quark in the two-point functions 
and further improvement is  
achieved by smearing also in the three-point functions, as described in 
the next section.

\section{New results for $B_0^s$ and $B_0^d$ mixing parameters}

%We have generated local and 1S smeared two-point functions at source and 
%sink together with three-point functions local at source and local and smeared 
%at sink, as defined in (\ref{corrdef}), in order to reduce the statistical 
%errors in our results. 

We have generated two-point functions with both local and 
smeared sources and sinks, using a smearing we call \emph{1S} since 
it takes an exponential form. Our three-point functions are local at the 
source (the site of the 4-quark operator), and have both local 
and smeared sinks at either end, with the same smearing as in the 
two-point case. This reduces  
the statistical+fitting error in our analysis. The general definition 
of these two-point and three-point functions is given in (\ref{corrdef}).

In addition to the matrix elements relevant in the determination of $B_0^s$ 
mixing parameters, we have also calculated those corresponding to 
$B_0^d$ mixing in full QCD. The $s$ and $b$ valence quarks masses are 
the physical ones, while the $d$ valence quark mass is the same as 
$m_d^{sea}$ for any ensemble. Two different lattice spacings have been 
studied, the MILC coarse lattice ($a=0.12$) and the MILC fine lattice 
(a=$0.09$). On the first one we have the correlation functions 
calculated for four different values of the light sea quark masses and 
on the second one, so far we have results only for one light sea quark 
mass. The parameters of the simulations, quark masses, number of 
configurations, number of time sources, etc, are shown in 
Table \ref{parameters}.

We have not analyzed yet the $1/M$ corrections for all the data 
collected in Table \ref{parameters}, so the results presented in these 
proceedings are only coming from the dominant contribution in the $1/M$ 
expansion. We are also still working on the fits with the lightest 
sea mass on the coarse ensemble, $m_q=0.005$, and results for this 
point will be presented elsewhere \cite{B0paper07}.

\subsection{Reduction of statistical+fitting errors} 

\begin{figure}
    \begin{center}              
 \includegraphics[width=9cm,height=12.5cm,angle=-90]{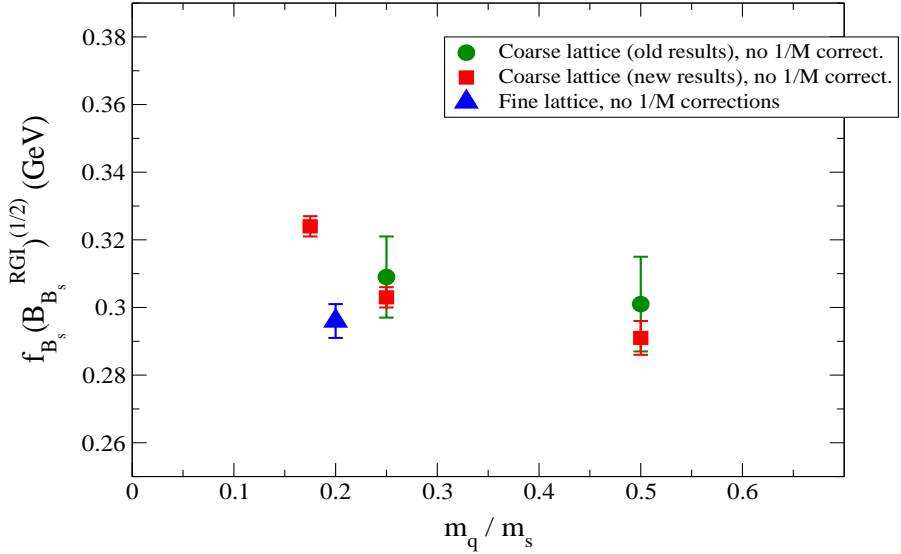} 
        \end{center}
\caption{${f_{B_s}\sqrt{\hat B_{B_s}}}(GeV)$. Errors are only statistical. 
\label{fBBsplot}}
\end{figure}

The use of several time sources and smearing greatly reduce the statistical 
errors as can be seen in Figure \ref{fBBsplot}. In that Figure, as an 
example of that reduction, we compare 
the results in our previous paper \cite{Bspaper} for 
${f_{B_s}\sqrt{\hat B_{B_s}}}(GeV)$ with our new results 
incorporating smeared correlation functions in the fits and new data. 
${f_{B_s}\sqrt{\hat B_{B_s}}}(GeV)$ is plotted as a function of the 
light sea quark mass over the physical strange quark mass, $m_q/m_s$,  
and the errors are only statistical.

With these new data we are able to get stable simultaneous fits for 
two-point and three-point correlation functions without any constraints 
in the two-point function parameters. With stable we mean that 
the central values, errors and $\chi^2/ndof$ do not change when 
we add more excited states in the functional forms to be fitted. 
The result is a reduction of statistical errors from 4.5\% to 1-2\% in 
${f_{B_s}\sqrt{\hat B_{B_s}}}(GeV)$, and similarly for 
$f_{B_d}\sqrt{\hat B_{B_d}}$. 

Another technique that could reduce further the size of statistical errors  
is the use of random wall sources for the light propagators. We have already 
checked that the statistical errors in the $B_s^0$ 
two-point parameters are improved by a factor of two, 
comparing results for the same heavy-light 
correlators we are using here but with HISQ \cite{hisq} 
(Highly Improved Staggered Quarks) 
light valence quarks, with and without random wall sources 
-see \cite{Kittalk} for more details about using random wall sources in 
heavy(NRQCD)-light(HISQ) correlators. Further study is needed to find 
how the use of this kind of source affect the three-point function 
parameters relevant for $B^0$ mixing.

\subsection{Calculation of the ratio $\xi$}

\begin{figure}
\begin{center}              
\includegraphics[width=9cm,height=12.5cm,angle=270]{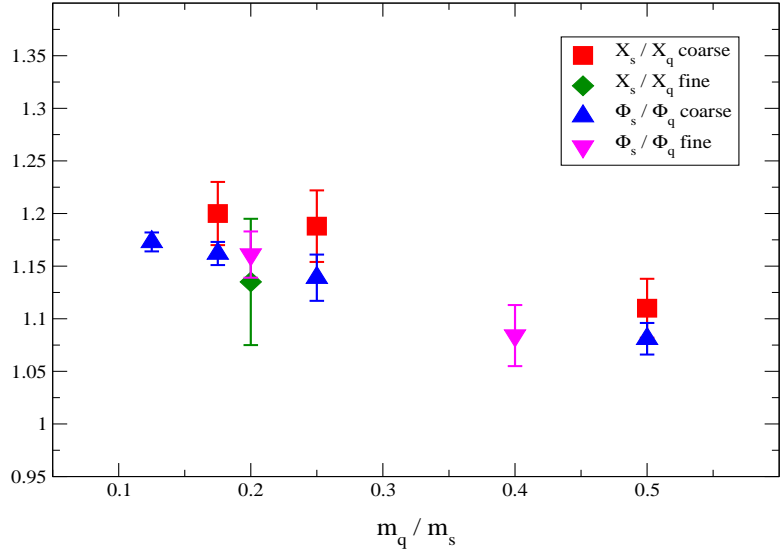} 
\end{center}
\caption{ $\frac{X_s}{X_q} = \frac{f_{B_s}\sqrt{B_{B_s}M_{B_s}}}
{f_{B_q}\sqrt{B_{B_q}M_{B_q}}}$ and $\Phi_s/\Phi_q=\frac{f_{B_s}\sqrt{M_{B_s}}}
{f_{B_q}\sqrt{M_{B_q}}}$ as a function of the light valence quark mass 
in the denominator. Errors are only statistical in all the quantities 
plotted. \label{xiplot}}
\end{figure}

Some of the errors affecting the calculation of $f_{B_q}\sqrt{B_{B_q}}$ 
will cancel almost completely and others partially in the ratio 
$\xi = \frac{f_{B_s}\sqrt{B_{B_s}}}{f_{B_d}\sqrt{B_{B_d}}}$. 
In Figure \ref{xiplot} we show values for this ratio multiplied 
by the square root of the masses of the $B_0^s$ and $B_0^d$ mesons, 
\ba
\frac{X_s}{X_q} = \frac{f_{B_s}\sqrt{B_{B_s}M_{B_s}}}
{f_{B_q}\sqrt{B_{B_q}M_{B_q}}}\,,
\ea 
together with the ratio 
$\Phi_s/\Phi_q=\frac{f_{B_s}\sqrt{M_{B_s}}}
{f_{B_q}\sqrt{M_{B_q}}}$, without the bag parameters from \cite{fBpaper}. 
The results are plotted as a function of $m_q/m_s^{phys.}$, where 
$m_q=m_d^{valence}=m_d^{sea}$. 

The errors for $X_s/X_q$ in Figure (\ref{xiplot}), which are 
only statistical, are larger than those for $\Phi_s/\Phi_q$ because 
we have not yet taken into account the correlations between 
the data in the numerator and denominator in this ratio. 
We expect to reduce this error to less than 2\% when these 
correlations are included (the current plotted  
values have $2.5\%$ errors). Another error that should be significantly 
reduced is that for the fine lattice point since we have not yet included 
all of our data.
        
The statistical errors are not the only ones to be reduced by taking 
the ratio. Discretization, relativistic and higher order operator 
matching will affect $f_{B_s}\sqrt{B_{B_s}}$ and $f_{B_d}\sqrt{B_{B_d}}$ 
in the same way and largely will cancel in the ratio. One expects their 
effects to come in at the level of the corresponding error in 
$f_{B_q}\sqrt{B_{B_q}}$ times $a(m_s-m_d)$ or $(m_s-m_d)/\Lambda_{QCD}$. The 
results for $f_{B_s}/f_{B_d}$ are nearly unchanged when adding one-loop  
and $1/M$ corrections \cite{fBpaper} and we expect something similar here. 
We have already checked that the difference between tree level and one-loop 
results is less than $1\%$. The scale $a^{-3}$ uncertainties, that 
lead to a $5\%$ error in $f_{B_q}^2B_{B_q}$, do not affect the ratio $\xi$.

The next step in our calculation will be to carry out a 
chiral extrapolation of these results to the physical 
point including the effect of taste-changing errors,  
to account for the remaining systematic in 
the calculation and remove the dominant light discretization errors.

\section{Summary and future work }

We have calculated the mixing parameters in the $B^0_s$ and $B^0_d$ 
systems for two different lattice spacings and five different light 
quark masses. The statistical errors have been reduced from our previous 
work by a factor of 2-3, so statistics is no longer a dominant source 
of uncertainty in the calculation of $f_{B_q}^2 B_{B_q}$. The largest 
error is now the uncertainty associated with the perturbative matching, 
that is also reduced from 9\% to 6.5\% by simulating on finer lattices. 
Further reduction of this source of error, as well as discretization 
errors, would also be possible by the use of MILC superfine lattices.

We also give preliminary results for the ratio $\xi$ 
versus $m_q/m_s$, where many theoretical uncertainties are partially 
or completely cancelled between denominator and numerator.

The analysis of ${1/M}$ corrections and results for ${m_d/m_s=0.125}$ 
and at least one other light quark mass on the fine lattice, will 
be presented in a forthcoming publication \cite{B0paper07}. We are also 
exploring different smearings and better fitting approaches to  
further reduce the statistical errors. In particular, we are 
getting promising preliminary results using random wall sources 
for the light propagators. 

Once other sources of errors have been reduced, we need to perform 
a chiral extrapolation of the $f_B\sqrt{B_B}$ and $\xi$ results 
incorporating light discretization uncertainties (taste-changing 
errors) and perturbative errors. We will also be able to 
perform a continuum extrapolation, since we have results for two 
different values of the lattice spacing.

Other talks on unquenched calculations of $B_0$ mixing parameters 
in this conference can be found in \cite{othertalks}.

\acknowledgments

This work was supported by the DOE and NSF (USA), by PPARC (UK) 
and by the Junta de Andaluc\'{\i}a [P05-FQM-437 and P06-TIC-02302] 
(E.G.). The numerical simulations were carried out 
at NERSC and Fermilab. We thank the MILC collaboration for use 
of their unquenched gauge configurations and the Fermilab collaboration 
for use of their asqtad propagators on the fine lattices.

\end{document}